# Roadmap FOR Establishing Interoperability of Heterogeneous Cellular Network Technologies -1-

**Hasni Neji, Ridha Bouallegue**

*Innov'Com Lab, Higher School of Communications of Tunis, Sup'Com*
*University of Carthage, Tunis, Tunisia.*

**Summary**
*The lack of interoperability between cellular access networks has long been a challenging burden, which telecommunication engineers and researchers are trying to overcome. In second generation networks for example, this problem lies in the lack of standardization. 3rd G networks is limited to a few operating modes using different radio transmission technologies that are not interoperable. 4G technology even being successful in its various trials cannot guarantee interoperability.*

*The undertaken approach to overcome this issue within heterogeneous networks begins by establishing a holistic understanding of cellular communication, and proposing an Ontological approach that expresses the domain's concepts, classes, and properties in a formal and unambiguous way. It begins by analyzing the structure of three different cellular technologies, and producing feature models. Lte-Advanced cellular network is the target of this ongoing analysis. The final objective sought is to build Ontology capable of providing a common view of cellular network technologies' domain.*

*Key words:*
*LTE-Advanced, Interoperability, Ontology, Feature modeling, UML.*

## 1. Introduction

This part of study presents a high-level description of Lte-Advanced features commingled with feature modeling. A feature is defined as a new or substantially enhanced functionality which represents added value to the system of belonging [1]. A feature should normally have an improved service to the network. Features are as independent as possible from each other, and relationships between features are explained in this setting of the mind or thoughts. The classification of features was biased by the future goal that we plan to achieve. Lte-Advanced network structure is described using a top-down approach: the network is logically divided in a number of sets, both from the architectural aspect and from the protocols aspect. From the architectural point of view, the sets are called "domains" (a domain is a group of entities). From the protocols point of view, the sets are called "strata" (a stratum is a group of protocols) [2]. These principles of networks description was introduced for first time in Universal Mobile Telecommunication Systems (UMTS). They do not correspond to any concrete realization in the network but were established mainly to organize and make the work easy].

## 2. Lte-Advanced network overview

LTE-Advanced (LTE-A) network technology was standardized by the 3rd Generation Partnership Project (3GPP), and endorsed by the International Telecommunications Union (ITU) as a fourth generation. The main objective of the LTE-A is to meet the challenge of increasing number of heterogeneous devices that requires higher bandwidth as well as heterogeneous networks. It is based on Orthogonal Frequency Division Multiplexing technology (OFDM). OFDM allows transmitting large amounts of digital data over a radio wave. It works by splitting the radio signal into multiple smaller sub-signals that are then transmitted simultaneously at different frequencies to the receiver. OFDM is a digital modulation technology in which in one time symbol waveform, more than thousands of orthogonal waves are multiplexed for increasing signal strength. OFDM provides some advantages. First of all, it offers a high robustness against selective fading. Secondly, allows for low-complexity implementation by means of Fast Fourier Transform (FFT) processing [3]. Finally, it offers spectrum flexibility which facilitates smooth evolution from already existing radio access technologies to LTE-Advanced.

## 3. LTE-Advanced Network architecture

### 3.1 The overall architecture

In the context of LTE-Advanced as a 4th Generation system, both the air interface (Radio-frequency portion of the circuit between the cellular phone set and the active base station ) and the radio access network are being enhanced or redefined. However, so far the core network architecture, i.e. the Evolved Packet Core (EPC), is not undergoing major changes from the already standardized System Architecture Evolution (SAE).

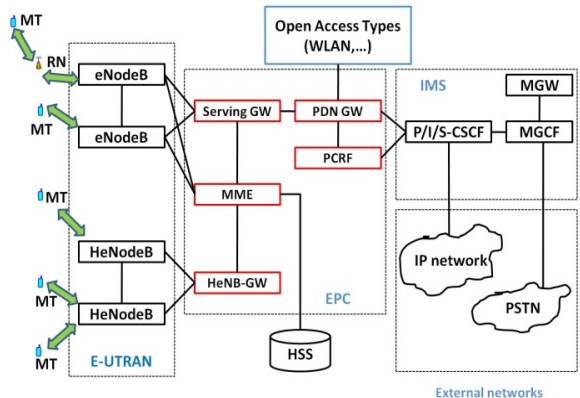

Figure 1. Long Term Evolution (LTE) Advanced Network Architecture (after [4]).

Fig.1 deals with the architecture of Long term evolution advanced [4]. WLAN stand for Wireless Local Area Network, PSTN: Public Switched Telephone Network, IP: Internet Protocol. The rest abbreviations of classes' names are introduced in detail in the essential LTE-Advanced feature list in table 1.

### 3.2 E-UTRAN architecture

The core part of the E-UTRAN architecture is the enhanced Node B, which provides the air interface with user plane and control plane protocol terminations towards the User Equipment. Fig. 2 is inferred from [5]. It shows the architecture of Evolved Universal Terrestrial Radio Access Network for LTE-Advanced. Each of the eNBs is a logical component that serves one or several E-UTRAN cells, and the interface interconnecting the eNBs is called the X2 interface. Moreover, Home eNBs (also called femtocells), can be connected to the EPC directly or via a gateway that provides additional support for a large number of HeNBs.  Further, 3GPP is considering relay nodes and sophisticated relaying strategies for network performance enhancement. eNBs provide the E-UTRAN with the necessary user and control plane termination protocols [5].

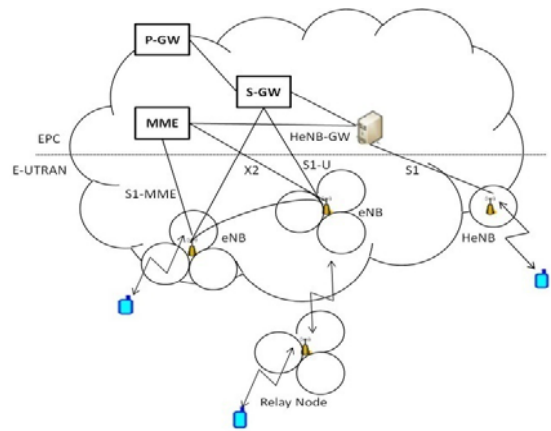

Figure 2. LTE-Advanced E-UTRAN architecture (from [5]).

The condensed graphical overview of both protocol strata in the user and control planes are presented in Fig. 3. In the user plane, the included protocols are the Packet Data Convergence Protocol (PDCP), the Radio Link Control (RLC), Medium Access Control (MAC), and Physical Layer (PHY) protocols. Whereas the control plane stack additionally includes the Radio Resource Control (RRC) protocols [5].

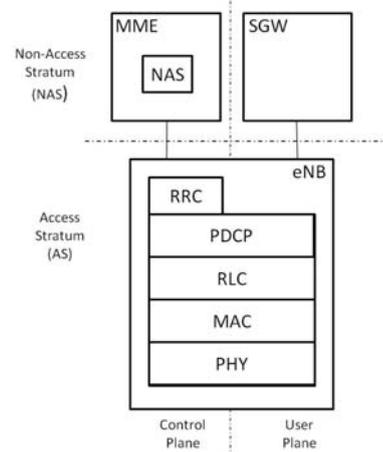

Figure 3. Protocol stack (after [5]).

### 3.3 The main protocol layers' functionalities

Here is a summary of the main functionalities carried out in each layer [5], [16], [17], [18], [19].
– PDCP (Packet Data Convergence Protocol)
- In-sequence delivery and retransmission of PDCP Session Data Units (SDUs) for acknowledge mode radio bearers at handover.

- Duplicate detection.
- Header compression.
- In-sequence delivery.
- Protocol error detection and recovery.
- Ciphering and integrity protection.
- RLC (Radio Link Control)
- Error correction through Automatic Repeat request (ARQ).
- Segmentation according to the size of the transport block and re-segmentation in case a retransmission is needed.
- Concatenation of SDUs for the same radio bearer.

– RRC (Radio Resource Control)
  - Mobility functions.
  - Quality of Service (QoS) management functions.
  - UE measurement reporting and control of the reporting.
  - Broadcast system information related to Non-Access Stratum (NAS) and Access Stratum (AS).
  - Security functions including key management.
  - NAS direct message transfer between UE and NAS.
  - Establishment, maintenance, and release of RRC connection.

– NAS (Non-Access Stratum)
  - Bearer context activation and deactivation.
  - Connection management between User Equipment (UE) and the core network.
  - Authentication.
  - Location registration management.
  - Registration.

– MAC (Medium Access Control)
  - Scheduling information reporting.
  - Padding.
  - Multiplexing and demultiplexing of RLC Packet Data Units (PDUs).
  - Local Channel Prioritization.

Error correction through Hybrid ARQ.

## 4 . Feature Modeling [6]

### 4.1 A quick idea about feature modeling

To perform an analysis of the Lte-Advanced cellular network technology, we proceed by producing a feature model for it. Attempting to define a feature model for this cellular network's technology allows us to explore, identify, and define the key concepts of it so that these aspects can be described in Ontology. It is this Ontology that then allows us to eventually improve interoperability between existing cellular network's technologies.

The feature model is an abstract representation of functionality found in the domain. It is used during domain engineering in order to obtain an abstract view on this functionality, which can be verified against the needs raised by the domain. Feature modeling is the activity of identifying essential characteristics of products in a domain and organizing them into a model called a feature model.

Product features are identified and classified in terms of capabilities, domain technologies, implementation techniques, and operating environments. Therefore, each feature is a relevant characteristic of the domain.

The description of feature models was tied to the introduction of the Feature-Oriented Domain Analysis (FODA-- Feature-oriented domain analysis (FODA) is a domain analysis method developed at the Software Engineering Institute (SEI). The method is known for the introduction of feature models and feature modeling--) approach in the late eighties. Czarnecki and Eisenecker [6] slightly modified and extended what was introduced in FODA (Features are typically arranged in a hierarchical structure that spans a tree) by adding some additional information, such as a short semantic description of each feature.

### 4.2 Lte-Advanced feature model

Defining a feature model for the Lte-Advanced network provides means to explore, identify, and define the key architectural aspects of this cellular network so that it would make it more explicit, comprehensible, and comparable with other networks. These aspects can then be described more fully in Ontology. It is this Ontology that can be used to establish interoperability between cellular communication network systems.

As shown in figure 4, the feature model is defined around concepts. The objective is to model features of elements and structures of a domain, not just objects in that domain. The silhouette of the network architecture (figure) was inserted to point the relative size and composition of the entire feature tree. For more detail on the mechanism of how to construct a feature model, the reader may consult [6], [7], or [8]. Figure 5 presents an excerpt of the LTE-Advanced feature model. It presents the LTE-Advanced architecture in a clear way via a feature tree.

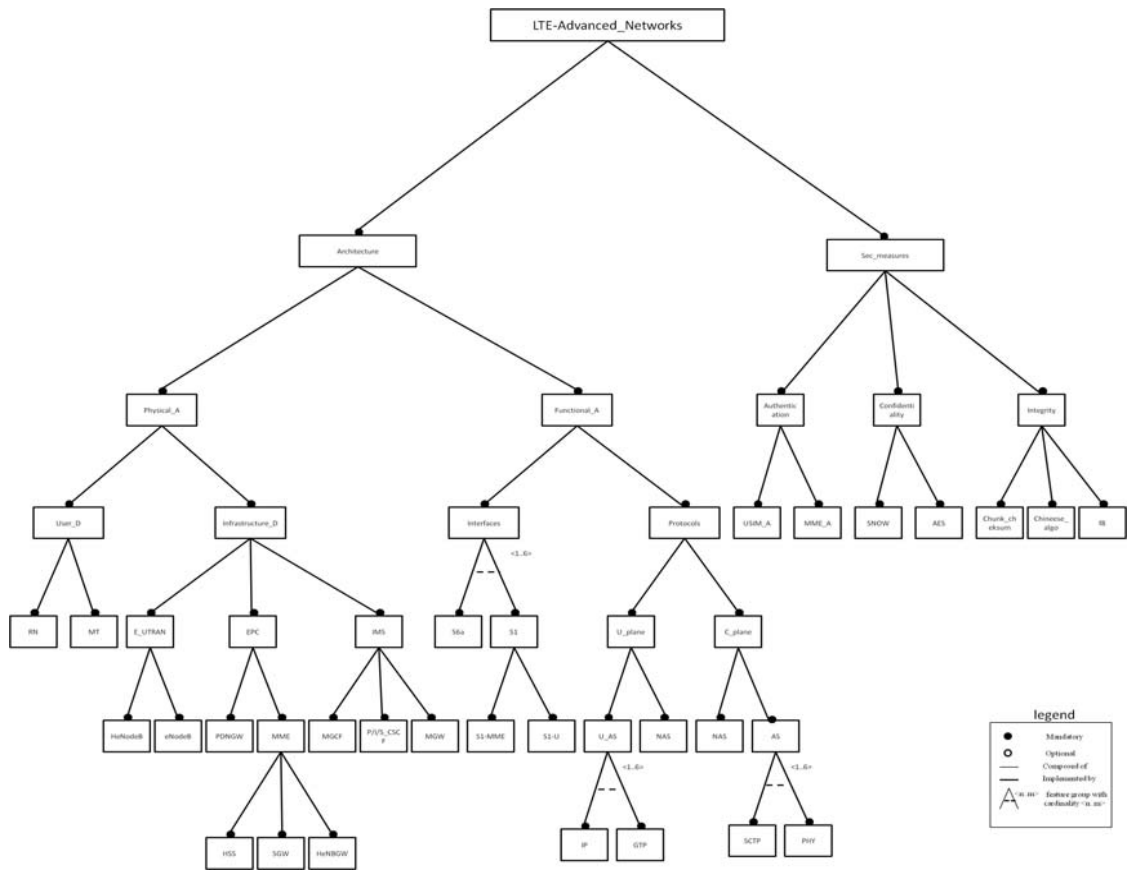

Figure 4. Feature model of LTE-Advanced network technology.

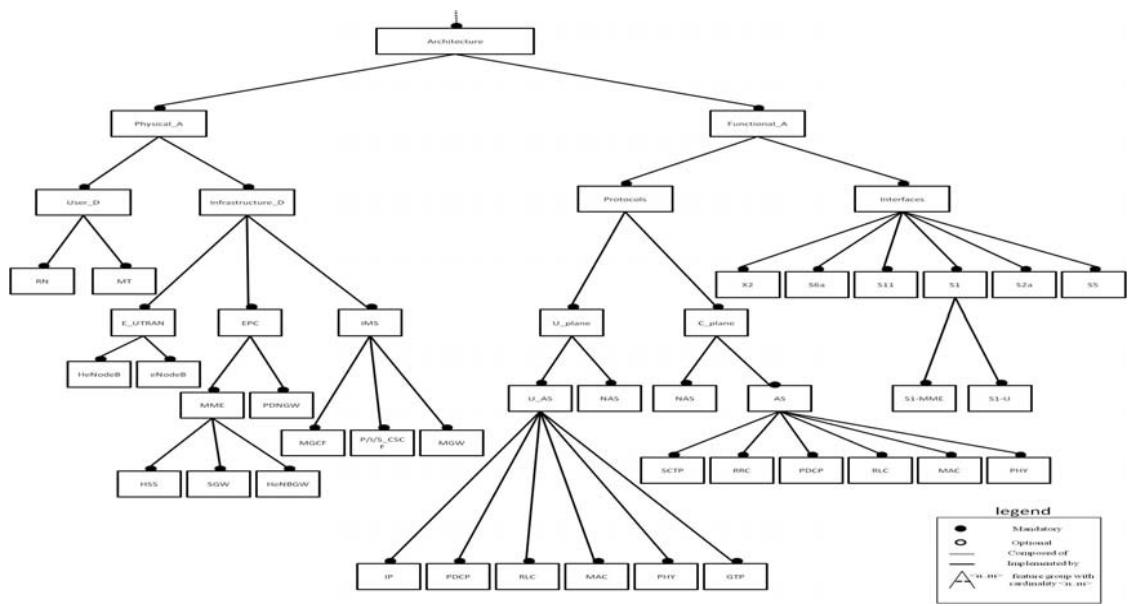

Figure 5. Excerpt of the LTE-Advanced feature model.

Figure 6 presents more specific piece of the LTE-Advanced feature model. The excerpt presents the LTE-Advanced physical architecture. It presents concepts inherited from both domains: user and infrastructure.

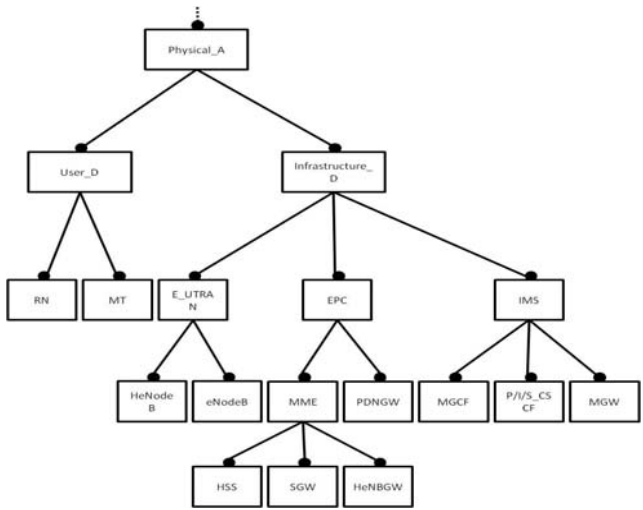

Figure 6. Excerpt of the LTE-Advanced feature model.

### 4.3 Methodology

[6] Provides an excellent methodology for gathering the information needed to construct a feature tree. He identifies the sources of features as the following:
• Domain experts,
• Existing systems,
• Pre-existing models (e.g., use-case models, object models…), and
• Models created during development process (i.e., features gotten during design).
Moreover, according to [6] the following set of general steps illustrates the feature modeling process:
• Record similarities between instances (i.e. common features).
• Record differences between instances (i.e. variable features).
• Analyze feature combinations and interactions.
• Record all the additional information regarding features.
• Organize the features in hierarchical feature tree with classification (mandatory, optional, alternative, and/or optional alternative features) [7].

✓ Mandatory. A mandatory child feature must be included in all the products in which its parent feature is included.
✓ Optional. An optional child feature can be optionally included in all products in which its parent feature appears.
✓ Alternative. A set of child features are defined as alternative, if only one of them can be selected when its parent feature is part of the product.

Or-relation. A set of child features are said to have an or-relation with their parent when one or more of them can be included in the products in which its parent feature appears.

### 4.4 Why considering feature modeling?

Feature modeling is one of the most popular domain analysis techniques, which analyzes commonality and variability in a domain in order to develop highly comprehensive and reusable core assets [9] for an interoperable system. However, feature modeling can be difficult and time-consuming without a precise understanding of its goals and the aid of practical guidelines.

The reason behind considering feature modeling is that feature-oriented domain analysis is an effective way to identify variability (and commonality) among different technologies in a domain. It is intuitive to express variability in terms of features. Feature modeling is considered a prerequisite for thorough understanding of the cellular communication networks technologies, and it is gaining popularity among practitioners and researchers. However, most users of feature modeling have difficulty in applying it to telecommunication engineering. Thus, in this paper, we strive to clarify what feature model is and how it is used, and provide practical usage of it to analyze the Lte-Advanced cellular network technology.

## 5. Essential LTE-Advanced network characteristics

The selected essential LTE-Advanced network characteristics that will be accounted for in building interoperability ontology for cellular network technologies are set forth in the following table (table 1).

Table.1 LTE-Advanced network Feature list.

| Feature | Full name of the feature | Description |
|---|---|---|
| Architecture | | |
| Physical_A | Physical Architecture | |
| User_D | User Domain | |
| RN | Relay Node | Relays are designed for coverage extension |
| MT | Mobile Termination | Mobile termination in the network of "B" is referred to when calls are routed to operator "B" via the network of operator "A", in order to be delivered to the end customer in the mobile network of "B". |
| Infrastructure_D | Infrastructure Domain | |
| E_UTRAN | Evolved Universal Terrestrial Radio Access Network | EUTRAN implements the LTE-Advanced access network as a network of eNBs. |
| HeNodeB | Home enhanced Node B | An evolved network component that serves one femtocell |
| eNodeB | enhanced Node B | An evolved network component that serves one cell. The eNB is responsible for many functions including: Radio Resource Management, IP header compression and user data encryption, the scheduling and allocation of uplink and downlink radio resources, and coordinating handover with neighboring eNBs. eNBs can communicate with multiple gateways for load sharing and redundancy. |
| EPC | Evolved Packet Core | LTE-Advanced's packet domain is a flat all-IP system designed for: much higher packet data rates, significantly lower-latency, the ability to optimize packet flows within all kinds of operational scenarios having to do with bandwidth rationing and charging schemes, explicit support for multiple radio access technologies in the interests of seamless mobility, and greater system capacity and performance. |
| PDNGW | Packet Data Network Gateway | Acts as a default router for the UE, and is responsible for anchoring the user plane for mobility between some 3GPP access systems and all non-3GPP access systems. |
| PCRF | Policy and charging rules functions | Is the single point of policy-based QoS control in the network. It is responsible for formulating policy rules from the technical details of Service Date Flows (SDF) that will apply to a user's services, and then passing these rules to the P-GW for enforcement. |
| MME | Mobility Management Entity | Supports many functions for managing mobiles and their sessions. |
| HeNBGW | H(e)NB Gateway | Is a mobile operator's equipment (usually physically located on mobile operator premises) through which the H(e)NB gets access to mobile operator's core network. HeNBGW serves as a concentrator for the C-Plane, specifically the S1-MME interface. The HeNBGW may optionally terminate the user plane towards the HeNB and towards the SGW, and provide a relay function for relaying User Plane data between the HeNB and the S-GW. The HeNB GW supports NAS |
| | | Node Selection Function (NNSF). |
| SGW | The Serving Gateway | Is responsible for anchoring the user plane for inter-eNB handover and inter-3GPP mobility. |
| HSS | Home Subscriber Server | Is the master database that stores subscription-related information to support call control and session management entities. |
| IMS | IP Multimedia Subsystem | It includes charging, billing and bandwidth management. |
| MGCF | Media Gateway Control Fonction | Communicates with the Call Session Control Function (CSCF) and controls the connections for media channels in an IMS-MGW. It performs protocol conversion between ISDN User Part (ISUP) and the IMS call-control protocols. |
| P/I/S_CSCF | proxy/interrogating/serving call session control function | Is the first contact point in IMS and interacts with GGSN (Gateway GPRS Support Node). |
| MGW | Media Gateway | It's a tool or service that transforms media flow between network telecommunication. |
| Functional_A | Functional_Architecture | |
| Interfaces | Interfaces | The LTE interfaces can be grouped into five categories: Air interface, E-UTRAN interfaces, Core network interfaces, Mobility and interworking interfaces, and service interfaces. |
| S6a | S6a | The Diameter-based reference point between the MME and the HSS. It enables transfer of subscription and authentication data. |
| S11 | S11 | Reference point between MME and SGW |
| S2a | S2a | The reference point between the PGW and any trusted access network that accommodates a user in the interests of seamless mobility. It provides the user plane with related control and mobility support between trusted non-3GPP IP access and the Gateway. S2a is based on Proxy Mobile IP. |
| S5 | S5 | The reference point between the SGW and the PGW. It provides user plane tunneling and tunnel management between SGW and PDNGW. It is used for SGW relocation due to UE mobility and if the SGW needs to connect to a non-collocated PDNGW for the required PDN connectivity. |
| X2 | X2 | The X2 interface is used to interconnect eNodeBs. It may be established between one eNodeB and some of its neighbor eNodeBs in order to exchange signalling information when needed. |
| S1 | S1 | The S1 interface connects the eNodeB to the EPC. It is split into two interfaces, one for the control plane and the other for the user plane. |
| S1-MME | S1-MME | the reference point for the control plane protocol between the E-UTRAN and the MME. |
| S1-U | S1-U | Reference point between EUTRAN and SGW for the per-bearer user plane tunneling and inter-eNB path switching during handover. |

| Protocols | Protocols | Protocols to define how different elements are able to communicate over the interfaces. |
|---|---|---|
| U_plane | User plane | The functions that deal with issues of user-to-user information transfer and associated controls such as flow control and error control mechanisms. |
| NAS | Non-access stratum | Is a functional layer between core network and user equipment. The layer supports signalling and traffic between those two elements. The NAS protocol performs authentication, registration, bearer context activation/ deactivation and location registration management |
| AS | Access stratum | Functional layer between radio network and user equipment. |
| IP | internet Protocol | Carry signal between eNodeB and PDN gateway. |
| PDCP | Packet Data Convergence Protocol | Packet Data Convergence Protocol): provides protocol transparency for higher-layer protocols, support for e.g., IPv4, and IPv6 (easy introduction of new higher layer protocols), and compression of control information (header compression). |
| RLC | Radio Link Control | RLC protocol provides logical link control over the radio interface. |
| MAC | Media Access Control | MAC protocol controls the access signalling (request and grant) procedures for the radio channel. |
| PHY | Physical layer | The physical layer implements OFDMA scheme on the downlink for high spectral efficiency, robustness against frequency-selectivity and multi-path interference. It supports flexible bandwidth deployment, facilitates frequency-domain scheduling and is well suited for Multiple Input Multiple Output (MIMO) techniques. |
| GTP | GPRS Tunneling Protocol | Is a group of IP-based communications protocols used to carry General Packet Radio Service (GPRS) within GSM, UMTS and LTE networks. |
| C_plane | Control plane | The functions defining all aspects of network signaling and control, such as call control and connection control. |
| NAS | Non-access stratum | Lie between the UE and MME. It executes functions and procedures that are completely independent of the access technology. These features include: authentication, security control procedures, idle mode mobility handling, idle mode paging procedures, charging and session management. |
| AS | Access stratum | Is a functional layer between radio network and user equipment. |
| SCTP | Stream Control Transmission Protocol | SCTP is a protocol unicast that allows the exchange of data in bidirectional mode between two end points SCTP. |
| RRC | Radio Resource Control | performs: Broadcast of System Information related to NAS and AS; Establishment, maintenance and release of RRC connection; Establishment, configuration, |
| | | maintenance and release of Signalling and Data Radio Bearers (SRBs and DRBs); Security functions including key management; Mobility functions including, e.g.: Control of UE cell selection/reselection; Paging; UE measurement configuration and reporting; Handover. |
| PDCP | Packet Data Convergence Protocol | The main functions of the PDCP for the control plane involve ciphering and integrity protection and transfer of control plane data. |
| RLC | Radio link control | Responsible for error recovery and flow control for data transportation between UE and eNodeB. The Radio Link Control (RLC) performs segmentation and concatenation to optimize the use of available resource, and tracks packets that were sent or received. |
| MAC | Medium Access Control | Responsible for logical-to-transport channel mapping, scheduling operation & random access. |
| PHY | Physical layer | The function is to provide data transport services on physical channels to the upper RLC and MAC sub-layers. |
| Sec_measures | Security measures | Provide the level of security required without impacting the user. |
| Authentication | Authentication | Authentication verifies the identity and validity of the SIM or USIM card to the network and ensures that the subscriber is authorized access to the network. |
| USIM_A | USIM_Authentication | In USIM authentication the f5 function must generate outputs before the f1 function. |
| MME_A | Mobility Management Entity-Authentication | This is a key control plane element. It is in charge of managing security functions (authentication, authorization, NAS signalling), handling idle state mobility, roaming, and handovers. |
| Confidentiality | Confidentiality | The property that information is not made available or disclosed to unauthorized individuals, entities or processes. |
| SNOW | SNOW | SNOW 3G is a stream cipher chosen by the 3rd Generation Partnership Project (3GPP) as a crypto-primitive to substitute KASUMI in case its security is compromised. |
| AES | Advanced Encryption Standard | The overall goal of the AES programme is to develop a Federal Information Processing Standard (FIPS) that specifies an encryption algorithm capable of protecting sensitive government information. |
| Integrity | Integrity | The property that data has not been altered in an unauthorized manner. |
| Chunk_cheksum | Chunk_cheksum | Scheme for SCTP, in which each data chunk contains its own checksum field to determine a corruption of data chunk. |
| f8 | f8 | KASUMI was the original algorithm for the f8 (data confidentiality) |

## 6. UML as an Ontology Description Language

Like feature models, a well-defined class diagram, part of the Unified Modeling Language (UML), can describe concepts and their relationships in a certain domain of discourse. Both approach have much to offer and can work together most effectively in an integrated environment. It is the reason behind their usage in this ongoing work.

The Unified Modeling Language (UML) is an attractive technique for modeling systems and documenting them. It uses symbols to graphically represent various components and their relationships within a system. UML may be described as the successor of object oriented design and analysis [10]. There are plenty of advantages behind its usage for development purposes, but the most important is to define some general purpose modeling language. UML diagrams are not only made for developers but also for users, and whoever interested in understanding a specific system. UML is not a development method rather it is joined with processes in purpose to make successful systems. To sum up, UML would be defined as a simple modeling mechanism to model all possible practical systems in complex environments.

As an important part of UML, class diagrams represent the object oriented view of a system in static way. They are generally used for development purposes. Moreover, they are widely relayed on at the time of system construction. Class diagrams (Fig. 7) show the existing classes of real world system, and how they are associated to each other in a hierarchical manner.

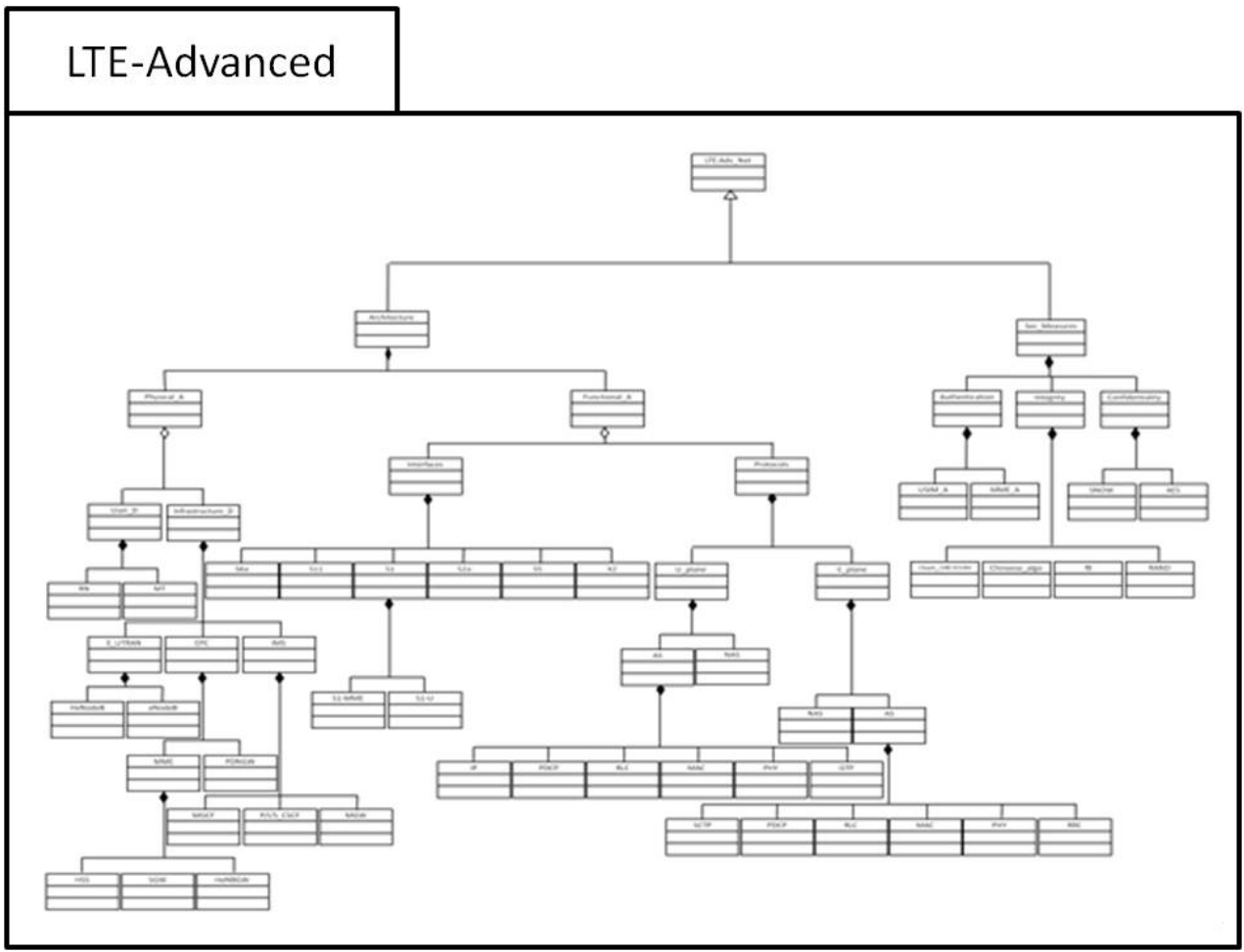

Figure 7. UML Description of LTE-Advanced network Features.

Fig. 7 illustrates a Package (the only one grouping thing available for gathering structural and behavioral things) of a global Lte-advanced class diagram. It is intended to show the relative size and composition of the entire ontology; it will be further shown in smaller diagrams (such as in Fig.8 representing high level classes) providing greater detail.

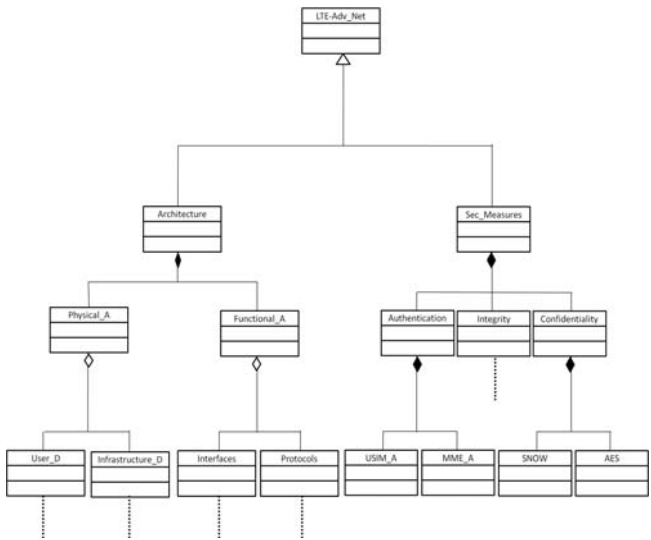

Figure 8. LTE-Advanced network class diagram.

As shown in fig. 8, each class is represented by a box with three parts: the first one at the top is for the name of the class, the second part is for the attributes of the class (specified by their name, type and visibility) and the third one is for the operations of the class (specified by name, argument list, return type and visibility).

For the purposes of annotating our ontologies and to keep figures light and clear, we do list neither the attributes nor the operations.

Note that full diamond is the symbole of Composition (Strongly-owned), meanwhile the empty one represents an Aggregation (non-strongly owned).

## 7. Contributions

The telecommunication contributions represented in this study are:
1. An initial investigation and analysis of the structure, and architecture of the Lte-Advanced cellular network technology, and the identification of essential characteristics of it.
2. The completion of a domain analysis of this technology and production of a feature model for it.
3. The preparation for an identification of the commonalities (from the point of view of architectural characteristics) between this network technology and other ones that must be accounted for in building a high level Ontology for the domain.
4. The use of feature models as a key asset to manage the commonalities and the variabilities of the cellular network's technologies. The common features then form the basis for the interoperability Ontology in the domain of interest.
5. The description of Lte-Advanced network structure using a top-down approach. The network was logically divided into domains and strata.
6. The use of UML to describe concepts and relationships in the cellular communication domain.

## 8. Conclusion

Developing an environment that maximizes interoperability, communication and efficiency tailored to particular domains is a common objective for researchers; who seek to improve the outputs in that domain. This development requires a flexible and trustworthy methodology. Feature modeling is the corner stone of it. The advantage of developing a specific feature model tailored to Lte-Advanced cellular network provides benefits stemming from representational efficiency. However, there has not been a lot of work in developing such analysis tailored to the domain of wireless cellular communication networks themselves. One reason for this is the amount of effort required to produce such feature diagram is substantial. Specific feature trees such as this ongoing one are, in fact, not easily buildable, which will require from us to undertake seemingly heavy processes to identify existing features to satisfy the representational needs.